\documentstyle[12pt]{article}
\newcommand{\ZJS}{\mbox{$Z_{S^{2}}(J)$}}

\newcommand{\yS}{\mbox{$\int_{S^{2}}d\mu$}}

\newcommand{\T}{\mbox{\rm Tr}}
\newcommand{\e}{\mbox{\rm exp}}

\newcommand{\TT}{\mbox{$t$}}

\newcommand{\MPL}{Mod. Phys. Lett. {\bf A}}
\newcommand{\NPB}{Nucl. Phys. {\bf B}}
\newcommand{\PRD}{Phys. Rev. {\bf D}}
\newcommand{\PLB}{Phys. Lett. {\bf B}}
\newcommand{\IJM}{Int. Jour. Mod. Phys. {\bf A}}
\newcommand{\A}{\mbox{$i$,$j$,$k$,$l$ all different}}
\newcommand{\B}{\mbox{$\delta^{ij}\varepsilon^{ik}\varepsilon^{il}
\varepsilon^{kl}$}}
\newcommand{\C}{\mbox{$\delta^{ij}\delta^{jk}\varepsilon^{il}$}}
\newcommand{\D}{\mbox{$i=j=k=l$}}
\newcommand{\E}{\mbox{$\delta^{ij}\delta^{kl}\varepsilon^{ik}$}}

\begin{document}

%%%%%%%%%%%%%%%%%%%%%% TITLE PAGE %%%%%%%%%%%%%%%%%%%%%%%%%%%%%%%%%%%%%%%
\begin{flushright}
hep-th/9510155 \\
BRX-TH-385     \\
October 1995
\end{flushright}
\begin{center}
{\Large{\bf The Master Field For 2D QCD On The Sphere}}

Jo\~ao P. Nunes\footnote{supported by JNICT, FLAD and Funda\c c\~ao C.
Gulbenkian (Portugal)}\\
and\\
Howard J. Schnitzer\footnote{supported in part by the DOE under grant
DE--FG02--92ER40706.}\\

Department of Physics\footnote{email address: NUNES,
SCHNITZER@BINAH.CC.BRANDEIS.EDU} \\
Brandeis University\\
Waltham, MA 02254

{\bf Abstract}
\end{center}

We continue our analysis of the field strength correlation functions of
two-dimensional QCD on Riemann surfaces by studying the large $N$ limit of
these correlation functions on the sphere for gauge group $U(N)$. Our results
allow us to exhibit an explicit master field for the field strength
$F_{\mu\nu}$ in a ``topological gauge'', given by a single master matrix
in the Lie algebra of the maximal torus of the gauge group. Field correlators
are obtained from traces of products of the master field. We also obtain a
master field for the gauge potential $A_{\mu}$ on the sphere, consistent with
the master field for the field strength.

%%%%%%% SECTION 1 INTRODUCTION%%%%%%%%%%%%%%%%%%%%%%%%%%%%%%%%%%%%%%%%%%%%
\section{Introduction}

Two dimensional Yang-Mills theories have provided an arena to test many of the
ideas related to non-perturbative features of gauge theories. Since
the exact partition function of the theory has been computed by a variety of
methods \cite{BR,EW,BT1,BT2}, a number of interesting issues may be studied.
It has been shown that the $1/N$ expansion of these theories allows them to be
formally represented as a string theory, in terms of sums of maps
from worldsheets to a target space $\Sigma_{g}$ \cite{DG}-\cite{PH}.
It was found by  Douglas and Kazakov
that the $U(N)$ theory on the sphere has a third-order phase transition at
large $N$ when $e^{2}(Area)=\pi^{2}$ \cite{DK}-\cite{RU}. Similar transitions
take place on $RP_{2}$ and for other gauge groups \cite{CNS}.

In \cite{JPNS} we were able to apply the beautiful path integral methods
developed by Blau and Thompson in \cite{BT1,BT2,BT4} to study the correlation
functions of the field strength
in 2d gauge theories over Riemann surfaces. We showed that these
correlators are essentially topological in the gauge appropriate to the
abelianization methods of Blau and Thompson. In this gauge the ``abelian''
components of the
electric field ($i.e.$ in the Lie algebra of the maximal torus $T$ of the gauge
 group $G$) carry the essential information about these correlation
functions.
Moreover, in the large $N$ limit on the sphere,
the correlators are dominated by the critical representation found in
\cite{DK}. We studied
\cite{JPNS} the large $N$ limit of the two, four and six-point
functions explicitly, and showed that these undergo a second-order phase
transition. In fact, all $2n$ point functions have a second-order phase
transition \cite{MCHSSN}, while of course all odd point functions vanish.

In this paper we will study the large $N$ limit of these correlation functions
on the sphere in more detail. As Douglas as remarked \cite{MDOU},
two-dimensional QCD on the Euclidean sphere is the case in two dimensions that
may be most representative of higher dimensions, as this is the only case where
the free energy behaves as $N^{2}$ for large $N$, as in higher dimensions, and
where a saddle point dominates the path integral. We will show that the field
strength correlators factorize at large $N$ as products of gauge invariant
correlators, as expected.
This, together with the consequences of the
abelianization of the theory, will enable us to compute the
master field
\cite{GG,MD} for the electric field on the sphere. This master field is
directly and simply related to the critical representation found by Douglas and
Kazakov \cite{DK}.
The master field for the field strength then enables us to compute the master
field for the gauge potential on the sphere. These master fields are given
in terms of a single master matrix in the Lie
algebra of the maximal torus of the gauge group.

In section 2 we present the field strength correlators and study the
two and four-point functions on the sphere in the large $N$ limit.
A general argument for the detailed structure of arbitrary correlators in the
large $N$ limit is presented.
In section 3 we exhibit the master field, and in
section 4 we close with some comments and conclusions.

%%%%%%%%%%%%%%%%%%%%% SECTION 2 %%%%%%%%%%%%%%%%%%%%%%%%%%%%%%%%%%%%%%%%%%%%

\section{The Correlators}

In \cite{JPNS} we considered the $U(N)$ gauge theory on a Riemann surface
$\Sigma_{g}$ coupled to a source $J(x)$, and computed the partition function
(here presented only for the sphere)
\begin{equation}
\ZJS=\int{\cal D}A_{\mu}\,\e[\frac{-1}{2e^{2}}\yS\,\T(\xi^{2})+2\yS\,\T(J\xi)]
\label{m2.1}
\end{equation}
where the scalar fields $\xi^{a}$$(x)$ are defined by
$F_{\mu\nu}(x)=\xi(x)\sqrt{g(x)}\epsilon_{\mu\nu}$, where $\xi=T^{a}\xi^{a}$
with $T^{a}$ a generator of the group, $d\mu=\sqrt{g(x)}d^{2}x$ the
riemannian measure for the metric $g_{\mu\nu}$ on $S^{2}$, and
$\epsilon_{\mu\nu}$ is the usual antisymmetric tensor with $\epsilon_{01}=1$.
Functional derivatives of (\ref{m2.1}) with respect to $J$ give the correlation
functions of the field strength. Naively, one might expect the correlators to
be trivial because (\ref{m2.1}) is gaussian in the field strength. However
this is not the case. The
correlators are non-trivial due to topological considerations.
Following the elegant path integral methods of Blau and Thompson
\cite{BT1,BT2}, one
writes the action in terms of an auxiliary scalar field which is then
conjugated to the
Lie algebra $t$ of the maximal torus $T$ of $U(N)$. This produces, in analogy
with the Weyl integral formula for integration of class functions on Lie
groups, a Weyl determinant. Integrating out the non-diagonal components of the
gauge field then produces an effective abelian theory for the $t$ components of
the gauge field. However, one must include all $T$-bundle topologies that were
generated by the choice of gauge, which is the origin of the non-trivial
behaviour of the electric field correlators \cite{JPNS}. The remaining
functional integral
over abelian fields can then be computed by means of the Nicolai map
\cite{BT1,BT2}. For a clear and detailed explanation of these issues we refer
the reader to \cite{BT1,BT2} as well as \cite{BT4}.
The final result is then \cite{JPNS},
\begin{eqnarray}
\ZJS=\sum_{l}\dim(l)^{2}\exp[\frac{-e^{2}A(l+\rho)^{2}}{2}]
\; \frac{1}{|W|}\sum_{\sigma\in W}\e\{2ie^{2}\yS\,[\sigma(l+\rho),J^{\TT}]\}
\label{m2.2}
\end{eqnarray}
where $A$ is the area of $S^{2}$, $l$ labels irreducible representations
of $U(N)$ and $\rho$ is the
half-sum of the positive roots of $su(N)$. Notice that here we will use a
nonstandard normalization of the partition function, writing $(l+\rho)^{2}$
instead of the usual $C_{2}(l)$ in the exponential in (\ref{m2.2}). This
just corresponds to
choosing the overall area dependent (but representation independent)
multiplicative factor in $Z_{S^{2}}$, and is an allowed renormalization choice
\cite{EW,BT2}.
This choice corresponds to the elimination of the contact term quadratic in
$J$ which
is present in the usual normalization \cite{JPNS}. Notice that this term was
responsible for the appearence of contact terms in the correlation functions in
\cite{JPNS}.
With our new choice of normalization, they
are renormalized away. Nevertheless we still have,
\begin{equation}
 \langle\T\xi^{2}(z)\rangle=2e^{2}\frac{d}{dA}F
\label{m2.3}
\end{equation}
relating the two-point function and the free energy $F$, as well as similar
results for higher derivatives of the free energy. Thus, no physics is changed
by our choice of normalization.

We can compute the electric field correlators from (\ref{m2.2}). As in
\cite{JPNS} all odd-point functions will vanish as a consequence of symmetry.
Also, notice that as an important consequence of abelianization (and
renormalization choices) only the $t$ components of the source $J$ enter in
(\ref{m2.2}),
so that only those components of $\xi$ lying in the Lie algebra of the maximal
torus $T$ will have nonvanishing correlators.
Therefore, in all that follows all Lie algebra indices inside correlation
functions will be in
$t$. Differentiating (\ref{m2.2}) with respect to $J$ at $J=0$
produces the 2n-point functions (normalized by $Z_{S^{2}}$),
\begin{equation}
\langle\xi^{a_{1}}(z_{1})\cdots\xi^{a_{2n}}(z_{2n})\rangle=(-1)^{n}
\frac{e^{4n}}{Z_{S^{2}}}
\sum_{l}\dim (l)^{2} \exp[\frac{-e^{2}A(l+\rho)^{2}}{2}]\;
\frac{1}{|W|}\sum_{\sigma \in
W}(l+\rho)^{\sigma_{a_{1}}}\cdots(l+\rho)^{\sigma_{a_{2n}}}
\label{m2.4}
\end{equation}
(Points on the sphere are denoted by $z_{i}$).
Notice that for $U(N)$, the Weyl group $W$ is just the symmetric group $S_{N}$
acting by permutation of the $N$ diagonal entries of elements in $t$.
For the two-point function we then have,
\begin{eqnarray}
\langle\xi^{a}(z_{1})\xi^{b}(z_{2})\rangle=
\frac{-e^{4}}{Z_{S^{2}}}\sum_{l}\dim(l)^{2}\e[\frac{-e^{2}A(l+\rho)^{2}}{2}]
[p^{ab}(l+\rho)^{2}+m^{ab}n^{2}]
\label{m2.5}
\end{eqnarray}
(Note we have the normalization
$v^{2}=\frac{1}{2}\sum_{i=0}^{N}v^{i}v^{i}$, for $v\in \TT$).
{}From \cite{JPNS}
\begin{equation}
\frac{1}{|W|}
\sum_{\sigma}(l+\rho)^{\sigma_{a}}(l+\rho)^{\sigma_{b}}=p^{ab}(l+\rho)^{2}+
m^{ab}n^{2}
\label{m2.6}
\end{equation}
where
\begin{eqnarray}
p^{ab}= \left\{ \begin{array}{cc}
                 \frac{-2}{N(N-1)}  & \mbox{if $a\neq b$} \\
                 \frac{2}{N}        & \mbox{if $a=b$}
                \end{array}
        \right.
\makebox[1in]{and}
m^{ab}= \left\{ \begin{array}{cc}
                 \frac{1}{N(N-1)}  & \mbox{if $a\neq b$} \\
                 0                 & \mbox{if $a=b$}
                \end{array}
         \right.
\label{m2.7}
\end{eqnarray}
and
$n=\sum_{i}l^{i}$ is the total number of boxes in the Young tableau, of row
lengths $l^{1}\geq l^{2}\geq\cdots\geq l^{N}$, defined by $l$.
In the large $N$ limit we can use continum variables,
$$
x=\frac{i}{N} \makebox[1.2in]{with} 0\leq x\leq 1
$$
\begin{eqnarray}
l(x)=\frac{l^{i}}{N} \makebox[1.3in]{and} \sum_{i}=N\int_{0}^{1}dx
\label{m2.8}
\end{eqnarray}
It is then useful to define,
\begin{equation}
h(x)=-l(x)+x-\frac{1}{2}=-(l+\rho)(x)
\label{m2.9}
\end{equation}
In this limit the partition function is dominated by the critical
representation $\bar{l}$, which solves the saddle point equation for the
effective action $S_{eff}(h)$. This is an integral equation for the density of
eigenvalues $u(h)=\frac{\partial x(h)}{\partial h}\;\leq 1$ (so that
$\int dx f(x)=\int dh u(h) f(h)$).
At weak coupling ($e^{2}Area< \pi^{2}$) this is
given by the semi-circle law, while for strong coupling ($e^{2}Area> \pi^{2}$)
one has more complicated expressions involving elliptic functions \cite{DK}.
In both phases of the theory,
the critical representation is self-conjugate, so that one has
$\bar{n}=n(\bar{l})=0$. Therefore, with $\lambda=e^{2}N$ held fixed in the
large $N$ limit, the two-point function becomes,
\begin{eqnarray}
\langle\xi^{i}(z_{1})\xi^{j}(z_{2})\rangle \longrightarrow
-\lambda^{2}\delta^{ij}\int_{0}^{1}{\bar h}^{2}(x)dx
\nonumber \\
\makebox[4in]{as $N\rightarrow \infty$}
\label{m2.2ptf}
\end{eqnarray}
We see that in the large $N$ limit only the diagonal term with $i=j$ case
gives a contribution, and
that this correlator is proportional to ${\rm Tr}{\bar h}^{2}=\int_{0}^{1}dx
{\bar h}^{2}(x)$.

The four-point function is,
%\newpage
\begin{eqnarray}
\langle\xi^{i}(z_{1})\xi^{j}(z_{2})\xi^{k}(z_{3})\xi^{l}(z_{4})\rangle=
\frac{e^{8}}{Z_{S^{2}}}\sum_{l}\dim(l)^{2}\e[\frac{-e^{2}A(l+\rho)^{2}}{2}
]\cdot \makebox[1.4in]{} \nonumber \\
\frac{1}{|W|}\sum_{\sigma}(l+\rho)^
{\sigma_{i}}(l+\rho)^{\sigma_{j}}(l+\rho)^{\sigma_{k}}(l+\rho)^{\sigma_{l}}
\makebox[0.5in]{}
\label{m2.10}
\end{eqnarray}
The Weyl sum is \cite{JPNS}
\begin{eqnarray}
\frac{1}{|W|}\sum_{\sigma}(l+\rho)^{\sigma_{i}}(l+\rho)^{\sigma_{j}}
(l+\rho)^{\sigma_{k}}(l+\rho)^{\sigma_{l}}=
\makebox[2in]{}
\nonumber \\
\{a^{ijkl}C_{4}(l)+
b^{ijkl}[(l+\rho)^{2}]^{2}+c^{ijkl}nC_{3}(l)+
d^{ijkl}n^{2}(l+\rho)^{2}+e^{ijkl}n^{4}\}
\label{m2.11}
\end{eqnarray}
where the coefficents of the various $W$ invariant terms are completely
symmetric in the indices $i$,$j$,$k$,$l$. We define Casimir operators by
\begin{equation}
C_{k}(l)=\sum_{i}[(l+\rho)^{i}]^{k} \makebox[2in]{for $k\geq 2$}
\label{m2.13}
\end{equation}
(Notice that the usual normalization for $k=2$ is
$C_{2}(l)=[(l+\rho)^{2}-\rho^{2}]$).
 With
\[ \varepsilon^{ij}= \left\{ \begin{array}{cc}
                             1 & \mbox{$i=j$} \\
                             0 & \mbox{$i\neq j$}
                             \end{array}
                      \right. \]
we have (with {\it no sum} on repeated indices in (\ref{m2.12})) \cite{JPNS}:
\begin{eqnarray}
a^{ijkl}= \left\{ \begin{array}{cc}
                  \frac{-6(N-4)!}{N!} & \A \\
                  \frac{2(N-3)!}{N!}  & \B \\
                  \frac{-(N-2)!}{N!}  & \C \\
                  \frac{-(N-2)!}{N!}  & \E \\
                  \frac{(N-1)!}{N!}   & \D
                  \end{array}
           \right.
\makebox[.3in]{}
b^{ijkl}= \left\{ \begin{array}{cc}
                  \frac{12(N-4)!}{N!} & \A \\
                  \frac{-4(N-3)!}{N!} & \B \\
                  0                   & \C \\
                  \frac{4(N-2)!}{N!}  & \E \\
                  0                   & \D
                  \end{array}
          \right.
\nonumber
\\
c^{ijkl}= \left\{ \begin{array}{cc}
                  \frac{8(N-4)!}{N!}  & \A \\
                  \frac{-2(N-3)!}{N!} & \B \\
                  \frac{(N-2)!}{N!}   & \C \\
                  0                   & \mbox{otherwise}
                  \end{array}
           \right.
\nonumber
\makebox[.3in]{}
d^{ijkl}= \left\{ \begin{array}{cc}
                  \frac{-12(N-4)!}{N!} & \A \\
                  \frac{2(N-3)!}{N!}   & \B \\
                  0                    & \mbox{otherwise}
                  \end{array}
            \right.
\\
e^{ijkl}= \left\{ \begin{array}{cc}
                  \frac{(N-4)!}{N!} & \A \\
                  0                 & \mbox{otherwise}
                  \end{array}
            \right.
\makebox[2in]{}
\label{m2.12}
\end{eqnarray}
Therefore the four-point function is given by
\begin{eqnarray}
\langle\xi^{i}(z_{1})\xi^{j}(z_{2})\xi^{k}(z_{3})\xi^{l}(z_{4})\rangle=
\frac{e^{8}}{Z_{S^{2}}}
\sum_{l}\dim(l)^{2}\e[\frac{-e^{2}A(l+\rho)^{2}}{2}]\cdot
   \makebox[1.3in]{} \nonumber \\
\{a^{ijkl}C_{4}(l)+b^{ijkl}\mbox{$[$}
(l+\rho)^{2}\mbox{$]$}^{2}
+c^{ijkl}nC_{3}(l)+d^{ijkl}n^{2}(l+\rho)^{2}+e^{ijkl}n^{4}\}
\makebox[0.2in]{}
\label{m2.14}
\end{eqnarray}
The large $N$ limit of the Casimir operators of (\ref{m2.13}) for
the critical representation are given by
\begin{equation}
C_{2k}({\bar l})=N^{2k+1}\int_{0}^{1}dx{\bar h}^{2k}(x)
\label{m2.15}
\end{equation}
(Odd degree Casimirs will vanish because ${\bar l}$ is self-conjugate).
 Therefore, in the large $N$ limit only the $a^{ijkl}$ and $b^{ijkl}$ terms in
(\ref{m2.14}) will contribute. Thus
\begin{eqnarray}
\langle\xi^{i}(z_{1})\xi^{j}(z_{2})\xi^{k}(z_{3})\xi^{l}(z_{4})\rangle
\longrightarrow \;
\lambda^{4}\{(\delta^{ij}\delta^{jk}\delta^{kl}\int_{0}^{1}dx{\bar h}^{4}(x))+
(\delta^{ij}\delta^{kl}\varepsilon^{ik}+{\rm permut.})(\int_{0}^{1}dx{\bar
h}^{2})^{2}\}     \nonumber \\
\makebox[4in]{as $N\rightarrow \infty$}
\label{m2.16}
\end{eqnarray}
where repeated indices are {\it not} summed. Contract equation (\ref{m2.16})
with
the generators
of the maximal torus, $E_{ii}/\sqrt{2}$ ($E_{ii}$ is the diagonal matrix with 1
in the ith position of the diagonal and zeroes everywhere else). The first term
in (\ref{m2.16}) produces $\langle{\rm Tr}\xi^{4}\rangle$, while the
second term will give $\langle{\rm Tr}\xi^{2}{\rm Tr}\xi^{2}\rangle$.
We will now argue that this structure holds for all higher-point functions. The
large $N$ limit correlators will vanish unless each $t$-index appears an
even number of times in the correlator, therefore
with no essential loss of information, we restrict ourselves to traced
correlators ($i.e.$ with all indices contracted with the generators of $T$, as
in the example above).

Let $k$ be an even integer, and consider
\begin{equation}
\frac{1}{N!}\sum_{\sigma}(l+\rho)^{\sigma_{i_{1}}}\cdots
(l+\rho)^{\sigma_{i_{k}}}
\label{m2.17}
\end{equation}
in the large $N$ limit. Equation (\ref{m2.15}) shows that the Casimirs behave
at large $N$ as $C_{r}\sim N^{r+1}$. Further
from (\ref{m2.4}), the correlation function of $k$ electric
fields has a factor of
$e^{2k}=\frac{\lambda^{k}}{N^{k}}$. Therefore, for large $N$, the
nonvanishing terms will be of the form
\begin{equation}
\langle\xi_{1}\cdots\xi_{k}\rangle\sim
\alpha\,\frac{\lambda^{k}}{N^{k}}C_{r_{1}}({\bar l})\cdots C_{r_{p}}({\bar l})
\label{m2.18}
\end{equation}
where $\sum_{i=1}^{p}r_{i}=k$, with the numerical coefficent $\alpha$ of
order $N^{-p}\sim \frac{(N-p)!}{N!}$. The Weyl group
average produces a coefficient of order $N^{-p}$, for $p$ distinct
Lie algebra indices in the set
$\{ i_{1},\dots ,i_{k}\}$ of indices in the
correlator $\langle\xi^{i_{1}}\cdots\xi^{i_{k}}\rangle$. Necessarily $p\leq N$.
 That is,
$\{ i_{1},\dots i_{k}\}$ is decomposed in $p$ groups, with $r_{i}$ elements
each, with the same $t$-index. On the
other hand, for such a set of indices, the Weyl group average produces a linear
combination of Casimirs of the critical representation ${\bar l}$, with total
degree $k$. Because of
the extra power of $N$ that each Casimir carries in (\ref{m2.15}),
only the term with $p$ Casimirs survives in the large $N$ limit. This will be
the term with the maximum number of Casimirs. Notice that in
this limit one must have each $r_{i}$ even for a non-vanishing result, since
${\bar l}$ is self-conjugate. Therefore the large $N$ correlators will vanish
unless the
same index appears an even number of times. When one contracts with the
generators of $T$, $E_{ii}/\sqrt{2}$, each one of the $p$ Casimirs can be
identified
as coming from a trace. The correlator will be the expectation value of
a product of
traces of powers of $\xi$, which will factorize in the expected way (see for
example \cite{early}) as the product of the correlators of the individual
traces, $i.e.$
\begin{equation}
\langle\frac{1}{N}{\rm Tr}\xi^{r_{1}}\cdots
\frac{1}{N}{\rm Tr}\xi^{r_{p}}\rangle=
\langle\frac{1}{N}{\rm Tr}\xi^{r_{1}}\rangle \cdots\langle\frac{1}{N}{\rm
Tr}\xi^{r_{p}}\rangle
\label{m2.19}
\end{equation}
Thus, one only needs to compute the expectation value of a trace of a power of
 $\xi$, $\langle\frac{1}{N}{\rm Tr}\xi^{r}\rangle$, which
we do in the next section.

%%%%%%%%%%% SECTION 3 %%%%%%%%%%%%%%%%%

\section{The Master Field}

The fact that only those elements of the field $\xi$ lying in $t$ contribute
to the
correlation functions makes it very easy to calculate traced correlators.
Since $E_{ii}E_{jj}=\delta^{ij}E_{ii}$,
\begin{equation}
\langle{\rm Tr}\xi(z_{1})\cdots \xi(z_{2k})\rangle=
\delta^{i_{1}i_{2}}\delta^{i_{2}i_{3}}\cdots \delta^{i_{2k-1}i_{2k}}\,2^{-k}
\,\langle\xi^{i_{1}}(z_{1})\cdots \xi^{i_{2k}}(z_{2k})\rangle
\label{m3.1}
\end{equation}
where there is {\it no summation} over repeated indices for the
$\delta^{ij}$'s.
That is, we must have all
indices equal $i_{1}=i_{2}=\cdots=i_{2k}$. In this case the Weyl group average
for the representation $l$ will give precisely a factor of $C_{2k}(l)$. Thus,
\begin{equation}
\langle{\rm Tr}\xi(z_{1})\cdots \xi(z_{2k})\rangle=\frac{1}{Z_{S^{2}}}
\frac{(-1)^{k}e^{4k}}{2^{k}}\sum_{l}\dim
(l)^{2}\exp[\frac{-e^{2}A(l+\rho)^{2}}{2}] C_{2k}(l)
\label{m3.2}
\end{equation}
In the large $N$ limit this becomes,
\begin{equation}
\frac{1}{N}\langle{\rm Tr}\xi(z_{1})\dots \xi(z_{2k})\rangle \longrightarrow
\frac{(-1)^{k}\lambda^{2k}}{2^{k}}\int_{0}^{1}dx{\bar h}^{2k}(x)
\label{m3.3}
\end{equation}
where (\ref{m2.15}) has been used.
Thus in the large $N$ limit, the traced correlators are given by the
traces of powers of ${\bar h}=-({\bar l}+\rho)$, here viewed as a matrix in
$t$, the Lie algebra of the maximal torus. Moreover, for correlation functions
 of products of $p$ traces,
upon contraction with the generators $E_{ii}/\sqrt{2}$, the term that survives
in the large $N$ limit is precisely the one given by the product of $p$
Casimirs
of the critical representation ${\bar l}$ in such a way that the expectation
value factorizes into the product of the expectation values of the individual
traces, as in (\ref{m2.19}).

We now have enough information to construct the master field for the field
strength. (For some early references see \cite{early}). Since in this gauge
all correlation functions are
independent of the positions where the fields $\xi$ are inserted, one needs
only {\it one} master matrix for $\xi$. The traces of powers
of the master matrix, ${\hat \xi}$, should reproduce the correlation functions
above, so that
\begin{equation}
\lim_{N \rightarrow \infty} \langle\frac{1}{N}{\rm Tr}\xi^{2k_{1}}\cdots
\frac{1}{N} {\rm Tr}\xi^{2k_{p}}\rangle=
\frac{1}{N}{\rm Tr}{\hat \xi}^{2k_{1}}\cdots \frac{1}{N}{\rm Tr}{\hat
\xi}^{2k_{p}}
\label{m3.4}
\end{equation}
However because of factorization and (\ref{m3.3}), this is given by
$$
(-1)^{k_{1}}\frac{\lambda^{2k_{1}}}{2^{k_{1}}}\int_{0}^{1}dx{\bar
h}^{2k_{1}}(x)\cdots
(-1)^{k_{p}}\frac{\lambda^{2k_{p}}}{2^{k_{p}}}\int_{0}^{1}dx{\bar
h}^{2k_{p}}(x)
$$
and therefore the master matrix ${\hat \xi}$ is given by
\begin{equation}
{\hat \xi}(x)=i\,\frac{\lambda}{\sqrt{2}}\,{\bar h}(x)
\makebox[2in]{for $0\leq x\leq 1$}
\label{m3.5}
\end{equation}
where $x$  is a diagonal Lie algebra
coordinate in the basis given by the $E_{ii}/\sqrt{2}$.
We see that as a remarkable consequence of abelianization, the master matrix
lies in $t$, the Lie algebra of the maximal torus $T$.

One can find a master field for the gauge field $A_{\mu}$, which satisfies
\\
$D_{\hat A}{\hat A}(z)={\hat F}(z)=\sqrt{g(z)}{\hat \xi}$. Since ${\hat \xi}$
is in $t$, this equation has solutions  with ${\hat A_{\mu}}$ also in $t$.
Such a solution cannot be globally defined on the
sphere as a consequence of the presence of the non-trivial $T$
bundles. (This can also be seen from the fact that ${\hat F}$ is proportional
to
the riemannian measure on the sphere). This is analogous to what happens with
the Dirac monopole. In fact,
for the usual riemannian measure on the sphere given by
$\sqrt{g}=\frac{A}{4\pi}\sin{\theta}$, ${\hat F}$ and ${\hat A}$ are
given by the solutions for the Dirac monopole tensored with the constant
matrix ${\hat \xi}$ in $t$, $i.e.$
\begin{eqnarray}
{\hat F}=\frac{A}{4\pi}\sin{\theta}\;d\theta \wedge d\phi \cdot {\hat \xi}
\makebox[2.5in]{} \nonumber \\
{\hat A}^{+}=\frac{A}{4\pi}(1-\cos{\theta})\;d\phi\cdot {\hat \xi}
\makebox[3in]{on the northern hemisphere}  \nonumber \\
{\hat A}^{-}=-\frac{A}{4\pi}(1+\cos{\theta})\;d\phi\cdot {\hat \xi}
\makebox[3in]{on the southern hemisphere}
\label{m4.1}
\end{eqnarray}
where $A$ is the area of the sphere.

The topological nature of the gauge invariant correlators is a consequence of
the invariance of the theory under area preserving diffeomorphisms. (For non
gauge-invariant correlation functions, the fact that they {\it are}
position independent is a consequence of the choice of gauge). This means that
the master field on the sphere can be chosen to be position independent, so
that one has one master matrix. This is also possible for the plane
\cite{GG}. Although we have only one master matrix and therefore we didn't use
the concept of free random variables of \cite{GG,MD,V}, one could still build a
Fock space representation for this master matrix as in \cite{GG,MD}.

%%%%%%%%%%%%%%% SECTION 4 %%%%%%%%%%%%%%%%%

\section{Conclusions}

In this paper we studied the large $N$ limit of the field strength correlation
functions for 2d QCD on the sphere. The use of the powerful abelianization
methods of \cite{BT2,BT4} provided us with a gauge where the correlators are
topological, and where essentially only the abelian components of the fields
contribute. This leads to a master field which is position independent, and
takes values in $t$, the Lie algebra of the maximal torus $T$ of the gauge
group. It
should be stressed that although the non-trivial $T$ bundles on $S^{2}$ give an
obstruction to the existence of a smooth global {\it diagonalizing} gauge
transformation map for $\xi$, the
{\it diagonalized} $\xi$ field exists globally on $S^{2}$ (see \cite{BT4}).
Therefore in this gauge, the master field for the field strength and the master
field for the gauge potential have the same geometry as the Dirac monopole.

A number of applications of the results presented in (\ref{m4.1}) are presently
under investigation.

{\bf Acknowledgements:}\\
We would like to thank M. Crescimanno, J. Isidro and H. Riggs for helpful
discussions.

%%%%%%%%%%%%%%%%%%%%%%%%%%%%%%%%%%%%%%%%%%%%%%%%%%%%%%%%%%%%%%%%%%%%%%%
%%%%%%%%%%%%%%%%%%% REFERENCES %%%%%%%%%%%%%%%%%%%%%%%%%%%%%%%%%%%%%%%%

\end{document}